\documentclass{article}
\usepackage{harvard}
\begin{document}

\title{The origin of the spacetime metric: Bell's `Lorentzian
  pedagogy' and its significance in general relativity\footnote{To
    appear in \emph{Physics meets Philosophy at the Planck Scale},
    C.~Callender and N.~Huggett (eds.), Cambridge University Press
    (2000).}}

\author{Harvey R. Brown\footnote{e-mail:
    harvey.brown@philosophy.oxford.ac.uk}\\Sub-Faculty of Philosophy,
  University of Oxford \and Oliver Pooley\footnote{e-mail:
    oliver.pooley@philosophy.oxford.ac.uk}\\Balliol College,
  University of Oxford}

\date{}

\maketitle

\begin{abstract}
  The purpose of this paper is to evaluate the `Lorentzian pedagogy'
  defended by J.S.~Bell in his essay ``How to teach special
  relativity'', and to explore its consistency with Einstein's
  thinking from 1905 to 1952. Some remarks are also made in this
  context on Weyl's philosophy of relativity and his 1918 gauge
  theory.  Finally, it is argued that the Lorentzian pedagogy---which
  stresses the important connection between kinematics and
  dynamics---clarifies the role of rods and clocks in general
  relativity.
\end{abstract}

\section{Introduction}\label{intro}

In 1976, J.S.~Bell published a paper on `How to teach special
relativity' \cite{bell76}.  The paper was reprinted a decade later in
his well-known book \emph{Speakable and unspeakable in quantum
  mechanics}---the only essay to stray significantly from the theme of
the title of the book.  In the paper Bell was at pains to defend a
dynamical treatment of length contraction and time dilation, following
``very much the approach of H.A.~Lorentz'' \cite[77]{bell87}.

Just how closely Bell stuck to Lorentz's thinking in this connection
is debatable. We shall return to this
question shortly.  In the meantime we shall briefly rehearse the
central points of Bell's rather unorthodox argument.

Bell considered a single atom modelled by an electron circling a more
massive nucleus, ignoring the back-effect of the field of the electron
on the nucleus.  The question he posed was: what is the prediction in
Maxwell's electrodynamics (taken to be valid relative to the
rest-frame of the nucleus) as to the effect on the electron orbit when
the nucleus is set (gently) in motion in the plane of the orbit? Using
only Maxwell's field equations, the Lorentz force law and the
relativistic formula linking the electron's momentum and its
velocity---which Bell attributed to Lorentz---he concluded that the
orbit undergoes the familiar longitudinal (``Fitzgerald'' (sic))
contraction, and its period changes by the familiar (``Larmor'')
dilation. Bell went on to demonstrate that there is a system of primed
variables such that the description of the moving atom with respect to
them coincides with that of the stationary atom relative to the
original variables, and the associated transformations of coordinates
is precisely the familiar Lorentz transformation.

Bell carefully qualified the significance of this result. He stressed
that the external forces involved in boosting a piece of matter must
be suitably constrained in order that the usual relativistic
kinematical effects such as length contraction be observed (see
section \ref{weyl} below). More importantly, Bell acknowledged that
Maxwell-Lorentz theory is incapable of accounting for the stability of
solid matter, starting with that of the very electronic orbit in his
atomic model; nor can it deal with cohesion of the nucleus. (He might
also have included here the cohesion of the electron itself.) How Bell
addressed this shortcoming of his model is important, and will be
discussed in section \ref{lp} below. In the meantime we note that the
positive point Bell wanted to make was about the wider nature of the
Lorentzian approach: that it differed from that of Einstein in 1905 in
both \emph{philosophy} and \emph{style}.

The difference in philosophy is well-known and Bell did not dwell on
it. It is simply that Lorentz believed in a preferred frame of
reference---the rest-frame of the ether---and Einstein did not,
regarding the notion as superfluous. The interesting distinction,
rather, was that of style. Bell argues first that ``we need not accept
Lorentz's philosophy to accept a Lorentzian pedagogy. Its special
merit is to drive home the lesson that the laws of physics in any
\emph{one} reference frame account for all physical phenomena,
including the observations of moving observers.'' He went on to stress
that Einstein postulates what Lorentz is attempting to prove (the
relativity principle). Bell has no ``reservation whatever about the
power and precision of Einstein's approach''; his point is that ``the
longer road [of FitzGerald, Lorentz and Poincar\'{e}] sometimes gives
more familiarity with the country'' \citeyear[77]{bell87}.

The point, then, is not the existence or otherwise of a preferred
frame---and we have \emph{no} wish to defend such an entity in this
paper. It is how best to understand, and teach, the origins of the
relativistic `kinematical' effects.  Near the end of his life, Bell
reiterated the point with more insistence:

\begin{quotation}
  ``If you are, for example, quite convinced of the second law of
  thermodynamics, of the increase of entropy, there are many things
  that you can get directly from the second law which are very
  difficult to get directly from a detailed study of the kinetic
  theory of gases, but you have no excuse for not looking at the
  kinetic theory of gases to see how the increase of entropy actually
  comes about. In the same way, although Einstein's theory of special
  relativity would lead you to expect the FitzGerald contraction, you
  are not excused from seeing how the detailed dynamics of the system
  also leads to the FitzGerald contraction.'' \cite[34]{bell92}
\end{quotation}

There is something almost uncanny in this exhortation. Bell did not
seem to be aware that just this distinction between thermodynamics and
the kinetic theory of gases was foremost in Einstein's mind when he
developed his fall-back strategy for the 1905 relativity paper (see section~\ref{chalk}).

It is the principal object of this paper to analyse the significance
of what Bell calls the `Lorentzian pedagogy' in both special
relativity and general relativity.  Its merit is to remind us that in
so far as rigid rods and clocks can be used to survey the metrical
structure of spacetime (and the extent to which they do will vary from
theory to theory), their status as \emph{structured} bodies---as
``moving atomic configurations'' in Einstein's words---must not be
overlooked.  The significance of the dynamical nature of rods and
clocks, and the more general theme of the entanglement between
kinematics and dynamics, are issues which in our opinion deserve more
attention in present-day discussions of the physical meaning of
spacetime structure.

\section{Chalk and cheese: Einstein on the status of special relativity theory}\label{chalk}

Comparing the explanation in special relativity (SR) of the \emph{non-null}
outcome of the celebrated 1851 Fizeau interferometry experiment---a
direct corroboration of the Fresnel drag coefficient---with the
earlier treatment given by Lorentz can seem like comparing chalk and
cheese.

From the perspective of SR the drag coefficient is essentially a
simple consequence (to first order) of the relativistic velocity
transformation law, itself a direct consequence of the Lorentz
transformations. The explanation appears to be entirely kinematical.
Lorentz, on the other hand, had provided a detailed dynamical
account---based on his theory of the electron---of the microstructure
of the moving transparent medium (water in the case of the Fizeau
experiment) and its interaction with the light passing through it
\cite{lorentz92}. In his 1917 text \emph{Relativity}, Einstein noted
with satisfaction that the explanation of Fizeau's experiment in SR is
achieved ``without the necessity of drawing on hypotheses as to the
physical nature of the liquid'' \cite[57]{einstein61}.

Yet Lorentz had achieved something remarkable, and Einstein knew it.
In deriving the drag coefficient from principles contained within his
theory of the electron, Lorentz was able to reconcile the null results
of \emph{first-order} ether-wind experiments (all of which
incorporated moving transparent media) with the claimed existence of
the luminiferous ether itself. There were few, if any, complaints that
such a surprising reconciliation was obtained on the basis of \emph{ad
  hoc} reasoning on Lorentz's part. But the case of the
\emph{second-order} ether-wind experiments was of course different,
and it is worth noting Einstein's take on these, again as expressed in
\emph{Relativity}.

In order to account for the null result of the 1887 Michelson-Morley
\mbox{(M-M)} experiment, Lorentz and FitzGerald assumed, says
Einstein, ``that the motion of the body [the Michelson interferometer]
relative to the aether produces a contraction of the body in the
direction of motion'' \citeyear[59]{einstein61}. Einstein's claim is not quite right.
In fact, both Lorentz and FitzGerald had correctly and independently
realised that it was sufficient to postulate any one of a certain
family of motion-induced distortions: the familiar longitudinal
contraction is merely a special case and not uniquely picked out by
the \mbox{M-M} null result.\footnote{For a recent account of the
  origins of length contraction see \citeasnoun{brown99}.} But this common
historical error (repeated by Bell) should not detain us, for the real
issue lies elsewhere. In SR, Einstein stresses, ``the contraction of
moving bodies follows from the two fundamental principles of the
theory [the relativity principle and the light postulate], without the
introduction of particular hypotheses '' \citeyear[59]{einstein61}.

The ``particular hypotheses'' of FitzGerald and Lorentz went beyond
the phenomenological claim concerning the distortion of rigid bodies
caused by motion through the ether. Both these physicists, again
independently, attempted to justify this startling claim by surmising,
not unreasonably, that the molecular forces in rigid bodies, and in
particular in the stone block on which the Michelson interferometer
was mounted, are affected by the ether-wind in a manner similar to
that in which electromagnetic interactions are so affected when viewed
from the ether rest-frame. Unlike their contemporary Larmor, neither
FitzGerald nor Lorentz was prepared to commit himself to the claim
that the molecular forces \emph{are} electromagnetic in origin. In
this sense, their courageous solution to the conundrum posed by the
\mbox{M-M} experiment did involve appeal to hypotheses outside what
Einstein referred to as the `Maxwell-Lorentz' theory.

Indeed, it was precisely their concern with rigid bodies that would
have made FitzGerald and Lorentz less that wholly persuaded by Bell's
construction above, as it stands. It is not just that Bell's atomic
model relies on post-1905 developments in physics. The point is rather
that Bell does not discuss the forces that glue the atoms
together---the analogue of the `molecular forces'---to form a rigid
body like Michelson's stone.\footnote{We are grateful to P.~Holland
  for emphasising this point (private communication).} (Bell of course
would have known that they were also electromagnetic in origin while,
as we have seen, FitzGerald and Lorentz were uncertain and unwilling
to commit themselves on this point.)

Returning to Einstein, in \emph{Relativity} he also mentioned the case
of predictions concerning the deflection of high-velocity electrons
(cathode- and beta-rays) in electromagnetic fields (1961, 56).
Lorentz's own predictions, which coincided with Einstein's, were
obtained by assuming \emph{inter alia} that the electron itself
deforms when in motion relative to the ether. It is worth recalling
that predictions conflicting with those of Lorentz and Einstein had
been made by several workers; those of M.~Abraham (an acknowledged
authority on Maxwellian electrodynamics) being based on the hypothesis
of the non-deformable electron.\footnote{Details of this episode can
  be found in \citeasnoun[chapter~1]{miller81}.} Einstein's point was that
whereas Lorentz's hypothesis is ``not justifiable by any
electrodynamical facts'' and hence seems extraneous, the predictions
in SR are obtained ``without requiring any special hypothesis
\emph{whatsoever} as to the structure and behaviour of the electron''
\citeyear[56, our emphasis]{einstein61}.

Whatsoever? Not quite, as we see in the next section. But for the
moment let us accept the thrust of Einstein's point. The explanations
of certain effects given by Maxwell-Lorentz theory and SR differ in
both style and degree of success: in some cases the Maxwell-Lorentz
theory actually seems incomplete. Yet there was a stronger reason for
the difference in style, and for the peculiarities of the approach
that Einstein adopted in 1905 (peculiarities which should be borne in
mind when evaluating claims---which resurface from time to time---that
Poincar\'{e} was the true father of SR).  Part of the further story
emerges in 1919, in a remarkable article Einstein wrote for the London
\emph{Times} \cite{einstein19}, when he characterised SR as an example
of a `principle theory', methodologically akin to thermodynamics, as
opposed to a `constructive theory', akin to the kinetic theory of
gases. Like all good distinctions, this one is hardly absolute, but it
is enlightening. It is worth dwelling on it momentarily.

In 1905 Einstein was faced with a state of confusion in theoretical
physics largely caused by Planck's 1900 solution to the vexing problem
of blackbody radiation. Not that the real implications of Planck's
quantum revolution were widely appreciated by 1905, even by Planck;
but that year saw Einstein himself publish a paper with the revolutionary
suggestion that free radiation itself had a quantised, or granular
structure \cite{einstein05a}. What his light-quantum proposal
undoubtedly implied in Einstein's mind was that the Maxwell-Lorentz
theory was probably only of approximate, or statistical validity. Now
within that theory, Lorentz, with the help of Poincar\'{e}, had
effectively derived the Lorentz (coordinate) transformations as the
relevant sub-group of the linear covariance group of Maxwell's
equations, consistent moreover with the FitzGerald-Lorentz deformation
hypothesis for rigid bodies.  But if Maxwell's field equations were
not to be considered fundamental, and furthermore the nature of the
various forces of cohesion within rigid bodies and clocks was obscure,
how was one to provide a rigorous derivation of these coordinate
transformations, which would determine the behaviour of moving rods
and clocks? Such a derivation was essential if one wanted, as Einstein
did, to tackle the difficult problem of solving Maxwell's equations in
the case of moving charge sources.

It is important to recognise that Einstein's solution to this
conundrum was the result of despair, as he admits in his
\emph{Autobiographical Notes} \cite{einstein69}.  Einstein could not see
any secure foundation for such a derivation on the basis of
``constructive efforts based on known facts'' \cite[53]{einstein69}.  In
the face of this \emph{impasse}, Einstein latched on to the example of
thermodynamics. If for some reason one is bereft of the means of
mechanically modelling the internal structure of the gas in a
single-piston heat engine, say, one can always fall back on the laws
of thermodynamics to shed light on the performance of that
engine---laws which stipulate nothing about the structure of the gas,
or rather hold whatever that structure might be. The laws or
principles of thermodynamics are phenomenological, the first two laws
being susceptible to formulation in terms of the impossibility of
certain types of perpetual motion machines. Could similar,
well-established phenomenological laws be found, Einstein asked, which
would constrain the behaviour of moving rods and clocks without the
need to know in detail what their internal dynamical structure is?

In a sense, Galileo's famous thought-experiment involving a ship in
uniform motion is an impossibility claim akin to the perpetual-motion
dictates of thermodynamics: no effect of the ship's motion is
detectable in experiments being performed in the ship's cabin. The
Galileo-Newton relativity principle was probably originally proposed
without any intention of restricting it to non-electromagnetic or
non-optical experiments \citeaffixed{brown&sypel95}{see}. In the light of the
null ether-wind experiments of the late 19th century, Einstein, like
Poincar\'{e}, adopted the principle in a form which simply restored it
to its original universal status. In Einstein's words:
\begin{quote}
  The universal principle of the special theory of relativity [the
  relativity principle] \ldots{} is a restricting principle for
  natural laws, comparable to the restricting principle of the
  non-existence of the \emph{perpetuum mobile} which underlies
  thermodynamics.  \cite[57]{einstein69}
\end{quote}

Turning to Einstein's second postulate, how apt, if at first sight
paradoxical, was its description by Pauli as the ``true essence of the
aether point of view'' \cite[5]{pauli81}. Einstein's light
postulate---the claim that relative to a certain `resting' coordinate
system, the two-way light-speed is constant (isotropic and independent
of the speed of the source)---captures that phenomenological aspect of
all ether theories of electromagnetism which Einstein was convinced
would survive the maelstrom of changes in physics that Planck had
started. Combined now with the relativity principle, it entailed the
invariance of the two-way light-speed.  This was not the only
application of the relativity principle in Einstein's 1905 derivation
of the Lorentz transformations \citeyear{einstein05b}, as we discuss in the next
section.\footnote{Nor were the two principles together strictly
  sufficient. The isotropy of space was also a crucial, if less
  prominent, principle in the derivation. Detailed examinations of the logic of Einstein's
  derivation can be found in \citeasnoun{brown&maia93} and
  \citeasnoun{brown93}. For an investigation of the far-reaching
  implications of abandoning the principle of spatial isotropy in the
  derivation, see \citeasnoun{budden97}.}

Einstein had now got what he wanted in the Kinematical Part of his
1905 paper, without committing himself therein to the strict validity
of Maxwell's equations and without speculation as to the detailed
nature of the cohesion forces within material bodies such as rods and
clocks. But there was a price to be paid. In comparing `principle
theories' such as thermodynamics with `constructive theories' such as
the kinetic theory of gases in his 1919 \emph{Times} article, Einstein
was quite explicit both that special relativity is a principle theory,
and that principle theories lose out to constructive theories in terms
of explanatory power:
\begin{quote}\ldots{} when we say we have succeeded in understanding a group
  of natural processes, we invariably mean that a constructive theory
  has been found which covers the processes in question.  \citeyear[228]{einstein82}\end{quote} This was essentially the point Bell was to make half
a century later.\footnote{In further elucidating the principle theory
  versus constructive theory distinction, we might consider the
  Casimir effect (attraction between conducting plates in the vacuum).
  This effect is normally explained on the basis of vacuum
  fluctuations in QED---the plates merely serving as boundary
  conditions for the q-number photon field.  But it can also be
  explained in terms of, say, Schwinger's source theory, which uses a
  c-number electromagnetic field generated by the sources in the
  plates so that the effect is ultimately due to interactions between
  the microscopic constituents of the plates.  (For a recent analysis
  of the Casimir effect, see \citeasnoun{rugh99}.) It might
  appear that the relationship between the first approach and the
  second is similar to that between Einstein's formulation of SR and
  the `Maxwell-Lorentz theory': the QED approach is simpler that
  Schwinger's and makes no claims as to the microscopic constitution
  of the plates (other than the claim that they conduct). But this
  appearance is misleading. Both approaches are equally `constructive'
  in Einstein's sense; it is just that one appeals to the quantum
  structure of the vacuum and the other to fluctuating dipole moments
  associated with atoms or molecules in the plates.}

\section{The significance of the Lorentzian pedagogy}\label{lp}

We saw in section \ref{intro} that Bell was aware in his 1976 essay of the
limitations of the Maxwell-Lorentz theory in accounting for stable
forms of material structure. He realised that a \emph{complete}
analysis of length contraction, say, in the spirit of the Lorentzian
pedagogy would also require reference to forces other than of
electromagnetic origin, and that the whole treatment would have to be
couched in a quantum framework. But it is noteworthy that Bell did not
seem to believe that \emph{articulation} of a \emph{complete}
dynamical treatment of this kind was a necessary part of the
Lorentzian pedagogy. In order to predict, on dynamical grounds, length
contraction for moving rods and time dilation for moving clocks, Bell
recognised that one need not know exactly how many distinct forces are
at work, nor have access to the detailed dynamics of all of these
interactions or the detailed micro-structure of individual rods and
clocks. It is enough, said Bell, to assume Lorentz covariance of the
complete dynamics---known or otherwise---involved in the cohesion of
matter. We might call this the \emph{truncated} Lorentzian pedagogy.

It is at this important point in Bell's essay that one sees something
like a re-run of the thinking that the young Pauli brought to bear on
the significance of relativistic kinematics in his acclaimed 1921
review article on relativity theory \cite{pauli81}. Pauli was struck by
the ``great value'' of the apparent fact that, unlike Lorentz,
Einstein in 1905 had given an account of his kinematics which was free
of assumptions about the constitution of matter. He wrote:

\begin{quotation}
  Should one, then, completely abandon any attempt to explain the
  Lorentz contraction atomistically? We think that the answer to this
  question should be No. The contraction of a measuring rod is not an
  elementary but a very complicated process. It would not take place
  except for the covariance with respect to the Lorentz group of the
  basic equations of electron theory, as well as of those laws, as yet
  unknown to us, which determine the cohesion of the electron itself.
  \cite[15]{pauli81}
\end{quotation}

Both Pauli and Bell seem then to contrast the dynamical underpinning
of relativistic kinematics with Einstein's 1905 argument. But it seems
to us that once the Lorentzian pedagogy relinquishes detailed
specification of the dynamical interactions involved---in other words
once it takes on the truncated form---the difference between it and
Einstein's approach, although significant, can easily be overstated. Indeed, we regard it as just
wrong to construe Einstein's 1905 `kinematical' derivation of the
Lorentz transformations as free of assumptions about the constitution
of matter, despite the distance between SR and the Maxwell-Lorentz
theory that Einstein urges (see above) in his \emph{Relativity}.

This is best seen in the second application of the relativity
principle in Einstein's argument. The first application, it will be
recalled, establishes the invariance of the two-way light-speed, given
the light postulate. Adopting the Einstein convention for
synchronising clocks in both the moving and rest frames, this entails
that the linear coordinate transformations take on the form of the
Lorentz transformations up to a non-trivial scale or conformal factor.
Einstein is now faced with the problem of reducing this factor to
unity (a problem which bedevilled Lorentz virtually throughout the
development of his theory of the electron). Einstein achieves this (as
did Poincar\'{e} independently) by a second appeal to the relativity
principle, in order to guarantee the group property of the
transformations---in particular to ground the claim that the form of
the transformation does not depend on the choice of frames.  This,
together with an appeal to the principle of spatial isotropy, does the
trick. The details need not concern us; the interesting question is
how this second application of the relativity principle should be
understood.

The coordinate transformations encode the behaviour of moving ideal
rulers and clocks under the crucial and universally accepted
assumption that these devices retain their rest lengths and periods
respectively under boosts. Suppose now that the coordinate transformations between
frames $S$ and $S^{\prime}$ are different in form from their inverse.
We expect in this case either the length contraction factor or the
time dilation factor (if any), or both, to differ when measured
relative to $S$ and when measured relative to $S^{\prime}$. And this
would imply a violation of the relativity principle. Specifically, it
would be inconsistent with the claim that the dynamics of all the
fundamental non-gravitational interactions which play a role in the
cohesion of these bodies satisfy the relativity principle. Thus the
\emph{dynamical} relativity principle constrains the form of the
\emph{kinematical} transformations, because such kinematics encodes
the universal dynamical behaviour of rods and clocks in motion.

It was clearly of importance to Bell that the Lorentzian pedagogy
relied on physics specified relative to a \emph{single} inertial frame
in order to account for the ``observations of moving observers'', and
in particular the very validity of the relativity principle itself.
But ultimately that physics amounted to the claim that the complete
theory of the construction of matter is Lorentz covariant, of which
the relativity principle \emph{inter alia} is a consequence. Einstein
on the other hand started with the relativity principle and the light
postulate, and derived (using the isotropy of space) Lorentz
covariance. In comparing these two approaches, two points must not be
lost sight of. The first is that Einstein's argument is dynamical,
since kinematics and dynamics in this context cannot be
disentangled.\footnote{The entangling of kinematics and dynamics is
  not peculiar to the relativistic context: details of a similar
  dynamical derivation of the Galilean transformations due to the
  fictional Albert Keinstein in 1705 are given in \citeasnoun{brown93}.} The
second point is that his `principle theory' approach to relativistic
kinematics ruled out the truncated Lorentzian pedagogy as a possible
starting point for Einstein.

\section{Einstein's unease about rods and clocks in special relativity}\label{rods}

The extent to which Einstein understood the full dynamical
implications of his 1905 derivation of the Lorentz transformations is
perhaps unclear.  Specifically it is not clear that he recognised the
role that rods and clocks can be seen to play in the derivation as
\emph{structured} bodies.  What is clearer is that he harboured, or
developed, a sense of unease about the status
of these bodies in his initial formulation of SR.  Einstein made use
of these devices in the first instance to operationalise the spatial
and temporal intervals, respectively, associated with inertial frames,
but he never explained where they come from. In an essay entitled
`Geometrie und Erfahrung' \cite{einstein21}, Einstein wrote:

\begin{quotation}
  It is \ldots{} clear that the solid body and the clock do not in the
  conceptual edifice of physics play the part of irreducible elements,
  but that of composite structures, which must not play any
  independent part in theoretical physics. But it is my conviction
  that in the present stage of development of theoretical physics
  these concepts must still be employed as independent concepts; for
  we are still far from possessing such certain knowledge of the
  theoretical principles of atomic structure as to be able to
  construct solid bodies and clocks theoretically from elementary
  concepts.  \cite[236, 237]{einstein82}
\end{quotation}

Einstein's unease is more clearly expressed in a similar passage in
his 1949 \emph{Autobiographical Notes}:

\begin{quotation}
  One is struck [by the fact] that the theory [of special relativity]
  \ldots{} introduces two kinds of physical things, i.e., (1)
  measuring rods and clocks, (2) all other things, e.g., the
  electromagnetic field, the material point, etc. This, in a certain
  sense, is inconsistent; strictly speaking measuring rods and clocks
  would have to be represented as solutions of the basic equations
  (objects consisting of moving atomic configurations), not, as it
  were, as theoretically self-sufficient entities. However, the
  procedure justifies itself because it was clear from the very
  beginning that the postulates of the theory are not strong enough to
  deduce from them sufficiently complete equations \ldots{} in order
  to base upon such a foundation a theory of measuring rods and
  clocks. \ldots{} But one must not legalize the mentioned sin so far
  as to imagine that intervals are physical entities of a special
  type, intrinsically different from other variables (`reducing
  physics to geometry', etc.). \citeyear[59, 61]{einstein69}
\end{quotation}

It might seem that the justification Einstein provides for the
self-confessed `sin' of treating rods and clocks as `irreducible', or
`self-sufficient' in 1905 is different in the two passages. In the
1921 essay, Einstein is saying that the constructive physics of atomic
aggregation is still too ill-defined to allow for the modelling of
such entities, whereas in the 1949 passage the point is that his 1905
postulates were insufficient in the first place to constrain the
theory of matter in the required way---which is little more than a
re-statement of the problem. But as we have seen, it was precisely the
uncertainties surrounding the basic constructive principles of matter
and radiation that led Einstein in 1905 to base his theory on simple,
phenomenological postulates.

Now there are two ways one might interpret these passages by Einstein.
One might take him to be expressing concern that his 1905 derivation
fails to recognise rods and clocks as complex, structured entities.
We argued in the previous section that this is not the case.  While
the derivation is independent of the details of the laws which
describe their internal structure, it is completely consistent with
the true status of rods and clocks as complex solutions of (perhaps
unknown) dynamical equations.  In fact, the derivation implicitly
treats them as such when the second appeal to the relativity principle
is made.

Alternatively, one might take Einstein to be concerned about the fact
that his postulates could not account for the availability of rods and
clocks in the world in the first place.  If this is his concern, then
it is worth noting that the possibility of the existence of rods and
clocks likewise does not follow from the mere assumption that all the
fundamental laws are Lorentz covariant.  It is only a full-blown
quantum theory of matter, capable of dealing with the formation of
stable macroscopic bodies that will fill the gap.

The significance of this point for the truncated Lorentzian pedagogy
and the constructive theory of rods and clocks in Einstein's thought
are themes we will return to in section~\ref{gr}. In the meantime, two points
are worth making.  First, it is perhaps odd that in his
\emph{Autobiographical Notes}, Einstein makes no mention of the
advances that had occurred since 1905 in the quantum theory of matter
and radiation. The understanding of the composition of bodies capable
of being used as rods and clocks was far less opaque in 1949 than it
was in 1905. Second, there is a hint at the end of the second passage
above that towards the end of his life, Einstein did not view
geometrical notions as fundamental in the special theory. An attempt
to justify this skepticism, at least in relation to four-dimensional
geometry, is given in section \ref{gr} below.

\section{A digression on rods and clocks in Weyl's 1918 unified field
  theory}\label{weyl}

In discussing the significance of the \mbox{M-M} experiment in his
text \emph{Raum-Zeit-Materie} \cite{weyl18a}, Hermann Weyl stressed that
the null result is a consequence of the fact that ``the interactions
of the cohesive forces of matter as well as the transmission of
light'' are consistent with the requirement of Lorentz covariance
\cite[173]{weyl52}. Weyl's emphasis on the role of ``the mechanics of
rigid bodies'' in this context indicates a clear understanding of the
dynamical underpinnings of relativistic kinematics. But Weyl's
awareness that rigid rods and clocks are structured dynamical entities
led him to the view that it is wrong to define the ``metric field'' in
SR on the basis of their behaviour.

Weyl's concern had to do with the problem of accelerated motion, or
with deviations from what he called ``quasi-stationary'' motion.
Weyl's opinion in \emph{Raum-Zeit-Materie} seems to have been that if
a clock, say, is undergoing non-inertial motion, then it is unclear in
SR whether the proper time read off by the clock is directly related
to the length of its world-line determined by the Minkowski metric.
For Weyl, clarification of this issue can only emerge ``when we have
built up a \textbf{dynamics} based on physical and mechanical laws''
\citeyear[177]{weyl52}. This theme was to re-emerge in Weyl's responses to
Einstein's criticisms of his 1918 attempt at a unified field theory.
Before turning to this development, it is worth looking at Weyl's
comments on SR.

In a sense Weyl was right. The claim that the length of a specified
segment of an arbitrary time-like curve in Minkowski
spacetime---obtained by integrating the Minkowski line element $ds$
along the segment---is related to proper time rests on the assumption
(now commonly dubbed the `clock hypothesis') that the performance of
the clock in question is unaffected by the acceleration it may be
undergoing. It is widely appreciated that this assumption is not a
consequence of Einstein's 1905 postulates. Its justification rests on
the contingent dynamical requirement that the external forces
accelerating the clock are small in relation to the internal
`restoring' forces at work inside the clock. (Similar considerations
also hold of course in the case of rigid bodies.)

Today we are more sanguine about the clock hypothesis than Weyl seems
to have been in \emph{Raum-Zeit-Materie}. There is experimental
confirmation of the hypothesis for nuclear clocks, for instance, with
accelerations of the order $10^8 \rm{cm/s}^2$.  But the question
remains as to whether the behaviour of rods and clocks captures the
full significance of the ``metric field'' of SR. Suppose accelerations
exist such that for no known clock is the hypothesis valid (assuming
the availability of the external forces in question!). Mathematically,
one can still determine---using the prescription above---the length of
the time-like worldline of any clock undergoing such acceleration if
it does not disintegrate completely. From the perspective of the
Lorentzian pedagogy, should one say that such a number has no physical
meaning in SR?  We return to this issue in the next section.

Weyl's separate publication of a stunning, though doomed unification of
gravitational and electromagnetic forces \cite{weyl18b} raised a number
of intriguing questions about the meaning of space-time structure
which arguably deserve more attention than they have received to date
\citeaffixed{ryckman94}{see, however,}.
Space prevents us from giving more than a sketch of the theory and its
ramifications; our emphasis will be on the role of the Lorentzian
pedagogy in evaluating the theory.

Weyl started from the claim that the pseudo-Riemannian space-time
geometry of Einstein's general relativity is not sufficiently local in
that it allows the comparison of the lengths of distant vectors.
Instead, Weyl insisted that the choice of unit of (spacetime) length
at \emph{each} point is arbitrary: only the ratios of the lengths of
vectors \emph{at the same point} and the angles between them can be
physically meaningful. Such information is invariant under a
\emph{gauge transformation} of the metric field: $g_{ij}\rightarrow
g_{ij}^{\prime}=e^{2\lambda(x)}g_{ij}$ and constitutes a conformal
geometry.

In addition to this conformal structure, Weyl postulated that
spacetime is equipped with an affine connection that preserves the
conformal structure under infinitesimal parallel transport. In other
words, the infinitesimal parallel transport of all vectors at $p$ to
$p^{\prime}$ is to produce a similar image at $p^{\prime}$ of the
vector space at $p$.\footnote{It is worth noting at this point that
  Weyl could, and perhaps should have gone further! As the keen-eyed
  Einstein was to point out, it is in the spirit of Weyl's original
  geometric intuition to allow for the relation between tangent spaces
  to be a weaker affine mapping: why insist that it be a similarity
  mapping?  Einstein made this point in a letter to Weyl in 1918. For
  details see \citeasnoun[102]{vizgin94}.} For a given choice of gauge, the
constant of proportionality of this similarity mapping will be fixed.
Weyl assumed that it differed infinitesimally from 1 and thereby
proceeded to show that the coefficients of the affine connection
depended on a one-form field $\phi_{i}$ in addition to the metric
coefficients $g_{ij}$ in such a way that the change in any length $l$
under parallel transport from $p$ (coordinates $\{x^{i}\}$) to
$p^{\prime}$ (coordinates $\{x^{i}+dx^{i}\}$) is given by:
\begin{equation}
dl=l\phi_{i}dx^{i}.\label{dlength}
\end{equation}
Under the gauge transformation $g_{ij}\rightarrow
g_{ij}^{\prime}=e^{2\lambda }g_{ij}$, $l\rightarrow e^{\lambda}l$.
Substituting this into (\ref{dlength}) gives:
\[
\phi_{i}\rightarrow\phi_{i}^{\prime}=\phi_{i}+\lambda_{,i},
\]
the familiar transformation law for the electromagnetic
four-potential. Weyl thus identified the gauge-invariant,
four-dimensional curl of the geometric quantity $\phi_{i}$ with the
familiar electromagnetic field tensor.

For a given choice of gauge a comparison of the length of vectors at
distant points can be effected by integrating (\ref{dlength}) along a
path connecting the points. This procedure will in general be path
independent just if the electromagnetic field tensor vanishes
everywhere.

As is well-known, despite his admiration for Weyl's theory, Einstein
was soon to spot a serious difficulty with the non-integrability of
length \cite{einstein18}. In the case of a static gravitational field, a
clock undergoing a round-trip in space during which it encountered a
spatially varying electromagnetic potential would return to its
starting point ticking at a rate different to that of a second clock
which had remained at the starting point and which was originally
ticking at the same rate.  An effect analogous to this `second clock
effect' would occur for the length of an infinitesimal rod under the
same circumstances. But it is a fact of the world---and a highly
fortunate one!---that the relative periods of clocks (and the relative
lengths of rods) do not depend on their relative histories in this
sense.

Before looking at Weyl's reply to this conundrum, it is worth
remarking that it was apparently only in 1983 that the question was
asked: what became of Einstein's objection once the gauge principle
found its natural home in quantum mechanics?  C.N.~Yang pointed out
that because the non-integrable scale factor in quantum mechanics
relates to phase, the second clock effect could be detected using
wavefunctions rather than clocks, essentially what Aharonov and Bohm
had discovered (Aharonov \& Bohm 1959; see also Ehrenberg \& Siday
1949).\footnote{\citeasnoun{yang84}.  Yang recounts this incident in \citeasnoun[18]{yang86}.} We note that Yang's question can be inverted: is there
a full analogue of the Aharonov-Bohm effect in Weyl's gauge theory?
The answer is yes, and it indicates that there was a further sting in
Einstein's objection to Weyl that he and his contemporaries failed to spot.
The point is that the second clock effect obtains in Weyl's theory
even when the electromagnetic field vanishes everywhere on the
trajectory of the clock, so long as the closed path of the clock
encloses some region in which there is a non-vanishing field.  This
circumstance highlights the difficulty one would face in providing a
dynamical or `constructive' account of the second clock effect in the
spirit of the full Lorentzian pedagogy.\footnote{It is worth noting
  that in 1923, Lorentz himself wrote, in relation to the rod analogue
  of the second clock effect in the Weyl theory, that this ``would
  amount to an action of an electromagnetic field widely different
  from anything that could reasonably be expected'' \cite[375]{lorentz37}.  But whether Lorentz was concerned with the dynamical problem
  of accounting for how Maxwell's electrodynamics could in principle
  have such an effect on physical bodies like rods---a consideration
  which one would not expect to be foreign to Lorentz' thinking!---or
  simply with the empirical fact that such an effect is non-existent,
  is not entirely clear from Lorentz's comments.}  Weyl's theory seems
to be bedevilled by non-locality of a very striking kind.

The precise nature of Weyl's response to Einstein's objection would
vary in the years following 1918 as he went on to develop new
formulations of his unified field theory based on the gauge principle
\citeaffixed{vizgin94}{see}.  But the common element was Weyl's rejection of the
view that the metric field could be assigned operational significance
in terms of the behaviour of rods and clocks. His initial argument was
an extension of the point he made about the behaviour of clocks in SR:
one cannot know how a clock will behave under accelerations and in the
presence of electromagnetic fields until a full dynamical modelling of
the clock under these circumstances is available. The price Weyl
ultimately paid for the beauty of his gauge principle---quite apart
from the complicated nature of his field equations---was the introduction of rather tentative
speculations concerning a complicated dynamical adjustment of rods and
clocks to the `world curvature' so as to avoid the second clock effect
and its analogue for rods.

We finish this section with a final observation on the nature of
Weyl's theory, with an eye to issues in standard general relativity to
be discussed shortly. We noted above that Weyl's connection is not a
metric connection. It is a function not only of the metric and its
first derivatives, but also depends on the electromagnetic gauge
field: in particular, for a fixed choice of gauge, the covariant
derivative of the metric does not vanish everywhere. What does this
imply?

The vanishing of the covariant derivative of the metric---the
condition of metric compatibility---is sometimes introduced
perfunctorily in texts on general relativity, but Schr\"{o}dinger was
right to call it ``momentous'' \cite[106]{schrodinger85}. It means that the local
Lorentz frames associated with a space-time point $p$ (those for
which, at $p$, the metric tensor takes the form ${\rm
  diag}(1,-1,-1,-1)$ and the first derivatives of all its components
vanish) are also local inertial frames (relative to which the
components of the connection vanish at $p$).\footnote{See
  \citeasnoun[313]{misner73}, 
\citeasnoun[16--20]{ehlers73} and \citeasnoun[57]{stewart91}.} If the laws of physics of the non-gravitational interactions
are assumed to take their standard special relativistic form at $p$
relative to such local Lorentz charts (the local validity of special
relativity), then metric compatibility
implies that gravity is not a force in the traditional sense---an
agency causing deviation from natural motion---, in so far as the
worldlines of freely falling bodies are geodesics of the connection.

The full physical implications of the non-metric compatible connection
in Weyl's theory remain obscure in the absence of a full-blown theory
of matter. Weyl's hints at a solution to the Einstein objection seem
to involve a violation of minimal coupling, i.e. a violation of the
prohibition of curvature coupling in the non-gravitational equations,
and hence of the local validity of special relativity. But it seems that the familiar
insight into the special nature of the gravitational interaction
provided by the strong equivalence principle---the encapsulation of
the considerations given in the previous paragraph---is lost in the
Weyl theory.

\section{The case of general relativity}\label{gr}

There is a recurrent, Helmholtzian theme in Einstein's writings
concerning Euclidean geometry: he claims that, as it is understood in
physics, it is the science ``des possibilit\'{e}s de deplacement des
corps solides'' \cite{einstein28}, or of ``the possibilities of laying
out or juxtaposing solid bodies'' (Einstein 1981, 163; see also
Einstein 1982, 234--235.).

But consider a universe consisting of some large number of mass points
interacting by way of the Newtonian gravitational potential. Few would
deny that a well-defined theory of such objects can be constructed
within the framework of Newtonian mechanics (or its recent Machian
counterparts such as \citeasnoun{barbourbertotti82}).  In such a world,
there is nothing remotely resembling rigid bodies or rulers which
allow for a direct operational significance to be assigned to
inter-particle distances. Yet these distances are taken to obey the
algebraic relationships of Euclidean geometry; either because this is a
foundational assumption of the theory (as in the Machian approach), or
because this is true of the particles' coordinate differences when
referred to the privileged coordinate systems with respect to which
the laws take on a canonical form. Moreover, the Euclidean constraint
on the instantaneous configuration of the particles is formally the
same as in a more ramified pre-general relativistic (quantum) theory
of matter which in principle allows for non-gravitational forces as
well, and hence for the possibility of the existence of stable, rigid
bodies.

Einstein was not oblivious to this point. He stressed that the
accepted theory of matter itself (even pre-special relativistic
theory) rules out the possibility of completely rigid bodies, and that
atomistic conceptions exclude ``the idea of sharply and statically
defined bounding surfaces of solid bodies.'' Einstein realised that
such physics ``must make use of geometry in the establishment of its
concepts'', concluding that ``the empirical content of geometry can be
stated and tested only in the framework of the whole of physics''
\cite[163]{einstein82}.

Rigid bodies furnish what is an already independently meaningful
spatial geometry with an (approximate) direct operational
significance, and the fact that they have this role is a consequence
of the theory of matter. It should be noted here that a necessary
condition for this state of affairs is that the dynamical equations
governing the non-gravitational interactions satisfy the so-called
Euclidean symmetries. But this more complicated and no doubt more
correct way of looking at things surely weakens the literal reading of
Einstein's original account of Euclidean geometry above as the science
of the possible arrangement of rigid bodies in space. The behaviour of
rigid bodies under displacements does not \emph{define} so much as
\emph{instantiate} the spatial geometry which might even have
primordial status in the foundations of the theory of such bodies. And
this point leads to another observation which is of considerable
relevance to this paper.

The fact that rods or rulers function as surveying devices for the
primordial Euclidean geometry is not because they `interact' with it;
the latter is not a dynamical player which couples to matter. In
arranging themselves in space, rigid bodies do not `feel' the
background geometry.  To put it another way, a rod is not a
`thermometer of space'. Nor is it in the intrinsic nature of such
bodies to survey space. It is the theory of matter which in principle
tells us what kind of entities, if any, can serve as accurate
geometrical surveying devices, and our best theories tell us rigid
bodies will do. One only has to consider the consequences of a
violation of the Euclidean symmetries in the laws of physics to dispel
any doubts in this connection. (All of this is to say that `what is a
ruler?' is as important a question in physics as `what is a clock?' An
answer to this last question ultimately depends on specifying very
special devices which `tick' in synchrony with an independently
meaningful temporal metric, a metric that might nonetheless be
specified by the {\em dynamics} of the total isolated system of which
the devices are a part.\footnote{See, for example, \citeasnoun[sections~3, 4 \& 12]{barbour94}.})

Turning now to the geometry of special relativity, what is of interest
is the behaviour of rods and clocks in relative motion. While the
Minkowski geometry does not play a primordial role in the dynamics of
such entities, analogous to the role which might be attributed to the
three-dimensional Euclidean geometry constraining relative distances,
it is definable in terms of the Lorentz covariance of the fundamental
dynamical laws.\footnote{A similar position is defended by \citeasnoun[326]{disalle95}. Our analysis of Euclidean geometry, and of the role of
  spacetime geometry in general relativity (see below) differs,
  however, from DiSalle's.} Hence spacetime geometry is equally not simply `the
science of the possible behaviour of physical rods and clocks in
motion'. All of the qualifications analogous to those we were forced
to consider in the case of Euclidean geometry apply. In the context of
SR, rods and clocks {\em are} surveying devices for a four-dimensional
geometric structure. But this is a structure defined in terms of the
symmetries of dynamical laws.  If matter and its interactions are
removed from the picture, Minkowski spacetime is not left
behind.\footnote{A more recent example in physics of an absolute
  geometrical structure with clear dynamical underpinnings is that of
  the projective Hilbert space (ray space) in quantum mechanics. A
  (curved) connection in this space can be defined for which the
  anholonomy associated with closed curves is the geometric phase of
  Aharonov and Anandan. This geometric phase encodes a universal
  (Hamiltonian-independent) feature of Schr\"{o}dinger evolution
  around each closed path in ray space, in a manner analogous to that
  in which the Minkowski geometry in SR encodes the universal
  behaviour of ideal rods and clocks arising out of the Lorentz
  covariant nature of the laws of physics.  For a review of geometric
  phase, see \citeasnoun{anandan92}; further comparisons with Minkowski
  geometry are spelt out in \citeasnoun{anandan91}.}  Rods and clocks do not
interact with a background metric field: they are not thermometers of
spacetime structure.

There is a temptation to extend this lesson to general relativity (GR)
in the following way. One might want to say that it is the local
validity of special relativity in GR---as defined at the end of the
previous section---that accounts for the existence of a metric field
of Lorentzian signature whose metric connection coincides with the
connection defining the inertial (free-fall) trajectories. The real
new dynamics in GR has to do not with the metrical properties of
spacetime but with the (generally) curved connection field and its
coupling to matter. In particular, a defender of the Lorentzian
pedagogy might be forgiven for accepting the maxim: no matter, no
metric (without, however, excluding the connection). As we have
argued, this is surely the right maxim for special relativity.  In a
universe entirely bereft of matter fields, even if one were to accept
the primordial existence of inertial frames, it is hard to attribute
any meaning in special relativity to the claim that empty spacetime
retains a Minkowski metric as an element of reality, or equivalently
to the claim that the inertial frames are related by Lorentz
transformations. (On this point Einstein seems to have been somewhat
inconsistent.  It is difficult to reconcile the remarks on special
relativity in his {\em Autobiographical Notes}, where he warns about
the reduction of physics to geometry, with the claim in Appendix~5 of
his {\em Relativity}---which he added to the fifteenth edition of the
book in 1952---that with the removal of all ``matter and field'' from
Minkowski space, this space would be left behind.\footnote{Actually,
  Einstein says that what remains behind is ``inertial-space or, more
  accurately this space together with the associated time'' , but
  subsequent remarks seem to indicate he meant the ``rigid
  four-dimensional space of special relativity'' \citeyear[171]{einstein82}.})

Adopting this position in GR has fairly drastic consequences in the
case of the vacuum (matter-free) solutions to Einstein's field
equations, of which `empty' Minkowski spacetime itself is one
solution. It entails that while the {\em flatness} of spacetime in
this case (and the curvature in other less trivial
solutions)---essentially an affine notion---may be said to have
physical meaning, the {\em metric} structure of the spacetime does not
(Brown 1997). (The metric might retain a meaning if one could adopt
Feynman's 1963 suggestion that such vacuum solutions are correctly
viewed as obtained by taking solutions involving sources and
matter-free regions and allowing these regions to become infinitely
large \cite[133--134]{feynman95}.  Feynman's view was
that gravitational fields without sources are probably unphysical,
akin to the popular view that all electrodynamical radiation comes
from charged sources.  Now the analogy with electromagnetism is
arguably not entirely happy given that gravity is itself a source of
gravity. Moreover, such an interpretation seems clearly inapplicable
to finite vacuum spacetimes.  Be that as it may, Feynman seemed to be
happy with the flat Minkowski vacuum solution, perhaps because he
could not entirely rid himself of his alternative `Venutian' view of
gravity as a massless spin-2 field on an absolute Minkowski
background.)

But the temptation to take the Lorentzian pedagogy this far should
perhaps be resisted. It overlooks the simple fact that the metric
field in GR (defined up to the diffeomorphic `gauge' freedom) appears
to be a {\em bona fide} dynamical player, on a par with, say, the
electromagnetic field. Even if one accepts---possibly either for
Machian reasons \citeaffixed{barbour94}{see} or with a view to quantum
gravity---the four-metric as less fundamental than the evolving curved
three-metrics of the Hamiltonian approach to GR, it is nonetheless
surely coherent to attribute a metric field to spacetime whether the
latter boasts matter fields or not. If absolute Euclidean distances
can exist in Newtonian universes bereft of rigid bodies, so much more
can the dynamical metric field in GR have claim to existence, even in
the non-generic case of the universal vacuum. Any alternative
interpretation of the metric field in GR would seem to require an
account of the coupling of a connection field to matter which was not
mediated by the metric field as it is in Einstein's field equations.
We know of no reason to be optimistic that this can be achieved.

Where does this leave the role of the Lorentzian pedagogy in GR? In
our opinion, it still plays a fundamental role in understanding in dynamical
detail how rods and clocks survey the metric field. To see this, let us consider
the following claim made by Torretti, in his magnificent 1983
foundational text on {\em Relativity and Geometry}. Torretti
formulates the basic assumption of GR as:

\begin{quote}
  The phenomena of gravitation and inertia \ldots{} are to be
  accounted for by a symmetric $(0,2)$ tensor field $\bf{g}$ on the
  spacetime manifold, which is linked to the distribution of matter by
  the Einstein field equations and defines a Minkowski inner product
  on each tangent space.  \citeyear[240]{torretti83}
\end{quote}

It follows immediately from this hypothesis that the Minkowski inner
product on tangent spaces induces a local approximate Minkowski
geometry on a small neighbourhood of each event.  Torretti claims that
one can thereby ``account for the Lorentz invariance of the laws of
nature referred to local Lorentz charts.'' The successes of special
relativity follow, says Torretti, from this local Minkowski geometry
\citeyear[240]{torretti83}.

In our view, this claim is a \emph{non sequitur}. It is mysterious to
us how the existence of a local approximate Minkowski geometry entails
the Lorentz covariance of the laws of the non-gravitational
interactions.  Theories postulating a Lorentzian metric but which
violate minimal coupling would involve non Lorentz covariant laws.
Equally, the primordial Euclidean geometry in the Newtonian theory of
mass points discussed at the start of this section does not entail
that the corresponding `laws of nature' (if there are any
non-gravitational interactions in the theory!) satisfy the Euclidean
symmetries. There is something missing in Torretti's account, and this
problem reminds one to some degree of the plea in some of
Gr\"{u}nbaum's writings for an account of why the ${\bf g}$ field has
the operational significance that it does. (A critical analysis of
Gr\"{u}nbaum's arguments, with detailed references, is given by
\citeasnoun[242--247]{torretti83}.)

It seems to us that the local validity of special relativity in GR
cannot be derived from what Torretti takes to be the central
hypothesis of GR above, but must be independently assumed. Indeed, it
often appears in texts as part of the strong equivalence principle,
taken as a postulate of GR \citeaffixed[386]{misner73}{for example,}. The assumption, which is intimately related to the postulate of
minimal coupling in GR, is that relative to the local Lorentz frames,
insofar as the effects of curvature (tidal forces) can be ignored, the
laws for the non-gravitational interactions take their familiar
special relativistic form; in particular, the laws are Lorentz
covariant. It is here of course that the full Lorentzian pedagogy can
in principle be used to infer the possibility of the existence of
material devices which more or less accurately survey the local
Minkowski geometry. In particular, it explains why ideal clocks, which
are chosen initially on dynamical grounds, act as hodometers, or
`way-wisers' of local Minkowski spacetime, and hence measure the
lengths of time-like curves over the extended regions in curved
spacetime that they traverse. (A very clear account of how the `hence'
in this statement was probably first understood by Einstein is given
by Torretti \citeyear[149--151]{torretti83}.  Elsewhere, Torretti \citeyear[312,
footnote~13]{torretti83} notes that as early as 1913 Einstein recognised that the
operational significance of the ${\bf g}$ field, and in particular the
significance of the null cones in the tangent spaces, required the
``separate postulate'' of the local validity of special relativity. It
seems from the above that Torretti did not wish to follow Einstein in
this respect.)

To conclude, the fact that general relativistic spacetimes are locally
Minkowskian only acquires its usual `chronometric' operational
significance because of the independent assumption concerning the
local validity of special relativity. Our main claim in this section
is that this point can only be understood correctly by an appeal to
the Lorentzian pedagogy.  Despite the fact that in GR one is led to
attribute an independent real existence to the metric field, the
general relativistic explanation of length contraction and time
dilation is simply the dynamical one we have urged in the context of
special relativity.

\section*{Acknowledgements}
We are grateful to the following people for helpful discussions
related to issues arising in this paper: J.~Armstrong,
G.~Bacciagaluppi, K.~Brading, P.~Holland, C.~Isham,
N.~Maxwell, S.~Saunders, D.~Wallace and particularly J.~Barbour. We are especially
indebted to J.~Butterfield for careful written comments on an earlier
draft.  Our thanks also go to the editors for the invitation to
contribute to this volume. One
of us (O.P.)\ acknowledges support from the UK Arts and Humanities
Research Board.

\end{document}